\documentclass[doc]{apa}
\usepackage{amssymb}
\usepackage{amsmath,amsthm,tabularx,curves,texdraw,psfrag}
\usepackage{apacite}
\usepackage[english]{babel}
\usepackage{lscape,dcolumn,hhline}
\usepackage[normalem]{ulem}
\usepackage{fancyhdr}
\usepackage{eurosym,bbding}
\usepackage{verbatim}
\usepackage{lineno}
\usepackage{placeins}
\usepackage{Sweave}
\usepackage{array}
\usepackage[caption=false]{subfig}
\usepackage{float}

\setkeys{Gin}{width=0.8\textwidth}

\title{Fitting Prediction Rule Ensembles to Psychological Research Data: An Introduction and Tutorial}
\shorttitle{Prediction Rule Ensembles}
\leftheader{Prediction Rule Ensembles}
\rightheader{Prediction Rule Ensembles}
\twoauthors{M. Fokkema}{C. Strobl}
\twoaffiliations{Universiteit Leiden}{Universit\"{a}t Z\"{u}rich}

\begin{document}

\maketitle

\section{Abstract}

Prediction rule ensembles (PREs) are a relatively new statistical learning method, which aim to strike a balance between predictive performance and interpretability. Starting from a decision tree ensemble, like a boosted tree ensemble or a random forest, PREs retain a small subset of tree nodes in the final predictive model. These nodes can be written as simple rules of the form \textit{if [condition] then [prediction]}. As a result, PREs are often much less complex than full decision tree ensembles, while they have been found to provide similar predictive performance in many situations. The current paper introduces the methodology and shows how PREs can be fitted using the \textbf{R} package \textbf{pre} through several real-data examples from psychological research. The examples also illustrate a number of features of package \textbf{pre} that may be particularly useful for applications in psychology: support for categorical, multivariate and count responses, application of (non-)negativity constraints, inclusion of confirmatory rules and standardized variable importance measures.\\[6pt]
\textit{Keywords}: R, recursive partitioning, decision making, machine learning\\

\noindent Supplementary Material available on \url{https://github.com/marjoleinF/met0000256/}.

\newpage

Statistical prediction problems are ubiquitous in psychology and related disciplines like sociology and medicine. In statistical prediction, a statistical model is fitted, in order to obtain a rule for predicting the value of the response or criterion variable, for observations for which only the values of the predictor variables are known. Such prediction problems are encountered in a wide range of areas in psychology: published examples include the prediction of children's social adjustment \cite<e.g.,>{Cric96}, violent behavior \cite<e.g.,>{YangyWong10}, workplace behavior \cite<e.g.,>{LeeyAlle02}, drinking behavior \cite<e.g.,>{LariyTurn04}, consumer attitudes \cite<e.g.,>{MaisGree04}, personality and personality disorders \cite<e.g.,>{KosiyWang16, ZanayFran06}, mood and anxiety disorders \cite<e.g.,>{GraayBijl02}, university dropout \cite{NiesyMeij16}, cardiovascular risks \cite<e.g.,>{HippyCoup08} and mortality \cite<e.g.,>{ChapyWeis16}.

With the increasing availability and sizes of datasets, there is an increasing interest in so-called statistical- or machine-learning methods, in which the central objective is to optimize predictive performance on observations not used in training the model \cite<e.g.,>{HarlyOswa16, HastyTibs09}. These methods have been contrasted with - and advocated over - more traditional explanatory statistical methods, which aim to develop a mechanistic model of the data-generating process that gave rise to the observed data \cite<e.g.,>{Brei01TwoCult, ChapyWeis16, YarkyWest17}. Often, these explanatory methods assume a parametric model, involving assumptions like normally distributed residuals or linear associations which may often be unrealistic in real-world data problems. Furthermore, in fields like genetics, bioinformatics and neuroimaging, datasets where the number of predictor variables far exceed the number of subjects are particularly common, but traditional parametric models are unable to deal with such data.

\subsection{Recursive Partitioning Methods}

Recursive partitioning methods (RPMs) provide a class of statistical learning methods that do not suffer from these disadvantages. RPMs like classification and regression trees, random forests and boosted tree ensembles have gained much popularity in recent years \cite<e.g.,>{MillyLubk16, StroyMall09}. These methods involve few if any assumptions about the data distribution and are able to deal with large numbers of potential predictor variables. RPMs recursively partition the space spanned by the predictor variables into increasingly smaller rectangular areas, in which the observations have increasingly similar response variable values. The resulting partition can be visualized as a decision tree; RPMs are therefore also known as decision tree methods. 

\begin{figure}[bt]
\setkeys{Gin}{width=0.5\textwidth}
\centering
\subfloat[Decision tree]{
\includegraphics{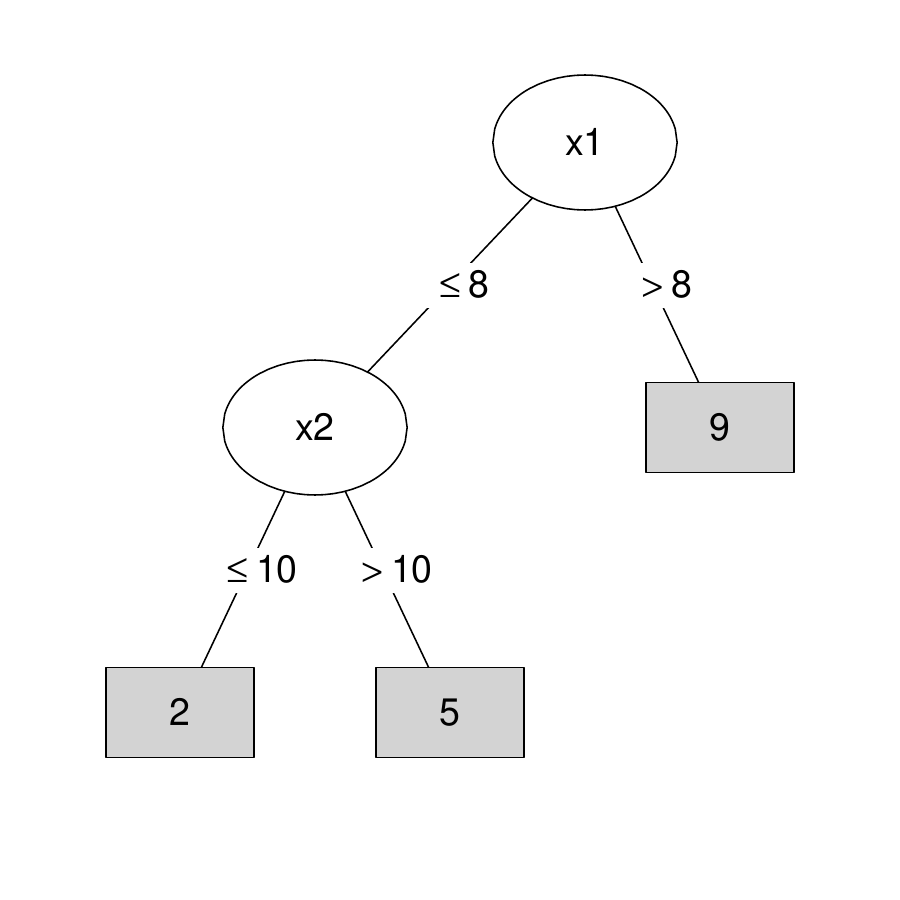}
}
\subfloat[Rectangular areas]{
\includegraphics{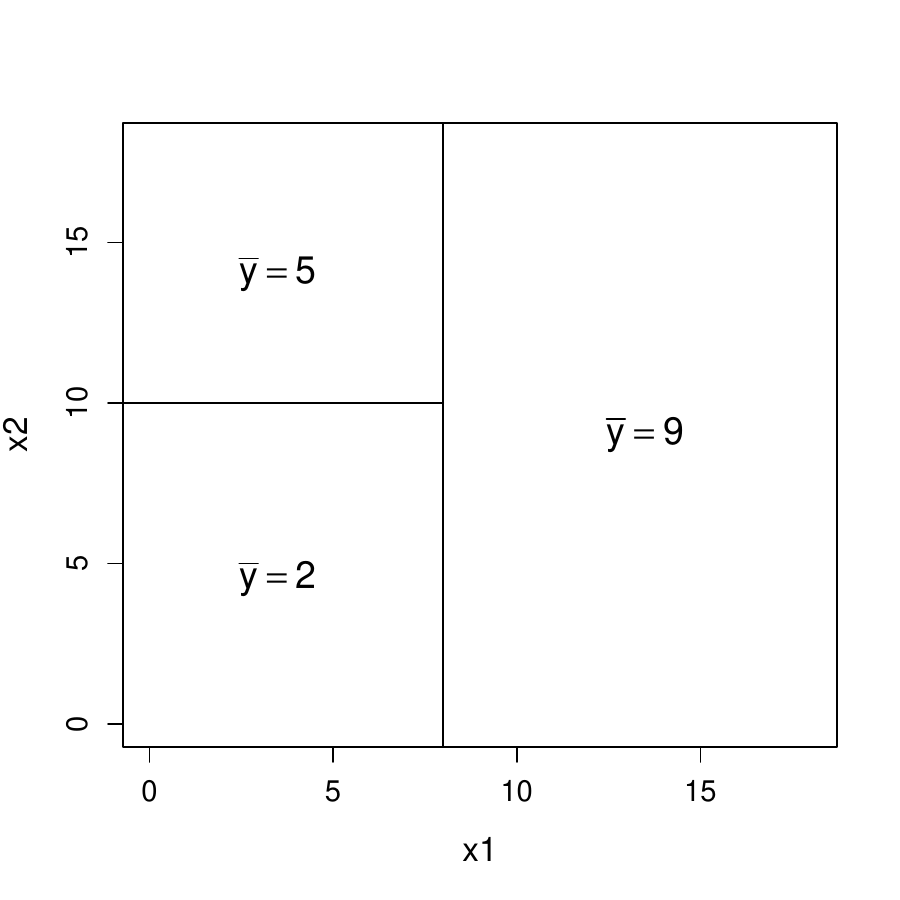}
}
\caption{\small Example recursive partition defined by two predictor variables ($x_1$ and $x_2$). In the left panel, the partition is depicted as a decision tree, with the node-specific means of the response variable in the terminal nodes. In the right panel, the partition is depicted as a set of rectangular areas, with the node-specific means of the response variable.}
\label{fig:exampletree}
\end{figure}

An example recursive partition or decision tree is depicted in Figure \ref{fig:exampletree}. The left panel shows the partition as a decision tree, of which the terminal nodes provide predictions for the value of the continuous response variable, based on the values of the predictor variables ($x_1$ and $x_2$). For new observations, for which the value of the response variable is not known but values of $x_1$ and $x_2$ are known, a prediction can be made by 'dropping' the new observations down the tree. The terminal node in which an observation ends up provides the prediction of the response. The right panel of Figure~\ref{fig:exampletree} shows how the decision tree separates the predictor variable space into rectangular areas, corresponding to the terminal nodes of the decision tree.

Decision trees like the one in Figure~\ref{fig:exampletree} involve only a single binary decision at every split to arrive at a final decision or prediction, which makes them relatively easy to interpret and apply. However, single decision trees provide lower predictive performance than other methods, and they can be unstable: small changes in the training data may yield large changes in the fitted tree \cite<e.g.,>{Brei98}. Ensemble techniques can improve on the predictive performance of unstable fitting procedures like single decision trees \cite{Brei96}. Techniques like bagging \cite{Brei96bagging}, boosting \cite{FreuyShap95} and random forests \cite{Brei01RandFor} grow a large number of trees on random samples of the data. Such tree ensemble methods have been shown to provide state-of-the-art predictive performance, exceeding that of any of the individual ensemble members \cite<e.g.,>{Roka10}. Introductions to tree ensemble methods specifically aimed at researchers in psychology can be found in \citeA{StroyMall09}, which discusses bagging and random forests, and \citeA{MillyLubk16}, which discusses boosted tree ensembles.  

A disadvantage of tree ensembles is that the improved predictive performance comes at the cost of reduced interpretability: instead of a single decision tree, the predictive model now consists of a large number of trees, that can no longer be visually grasped. To aid in interpretation of tree ensembles, variable importance measures have been developed, which aim to quantify the influence of individual variables on the predictions of the ensemble \cite<e.g.,>{Brei01RandFor, Frie01}. However, these importance measures present with several difficulties of their own, including a bias towards variables with a larger number of unique values and inflation under multicollinearity \cite<e.g.,>{AltmyTolo10, NicoyMall10, StroyBoul08, StroyBoul07}.

\FloatBarrier

\subsection{Prediction Rule Ensembles}

To reconcile the aims of predictive performance and interpretability, several authors have proposed prediction rule ensemble (PRE) methods \cite<e.g.,>{DembyKotl10, FrieyPope08, Mein10}. PREs consist of a small set of prediction rules: statements of the form \textit{if [condition] then [prediction]}. In fact, the regression tree in Figure~\ref{fig:exampletree} can also be written as a PRE, as shown in Table~\ref{tab:tree_as_pre_1}. The terminal node in the middle of the tree in Figure~\ref{fig:exampletree} is taken as the reference category, so that the intercept in Table~\ref{tab:tree_as_pre_1} corresponds to the mean in this terminal node. The two rules code membership of the remaining terminal nodes: the rules are 0-1 coded dummy variables, taking a value of 1 is their conditions apply and 0 if not. Thus, membership of the left-most node in Figure~\ref{fig:exampletree} is coded by the first rule in Table~\ref{tab:tree_as_pre_1}; the mean for that node is reproduced by summing the intercept and the first rule's coefficient (i.e., $5 - 3 = 2$). Similarly, the mean of the right-most terminal node is reproduced by summing the intercept and the second rule's coefficient (i.e., $5 + 4 = 9$).

\begin{table}[b]
\centering
\begin{tabular}{lc}
\hline
Function & Coefficient \\
\hline
Intercept & 5 \\
$I(x_1 \leq 8 \: \& \: x_2 \leq 10)$ & -3 \\
$I(x_1 > 8)$ & 4 \\
\hline
\end{tabular}
\caption{The decision tree in Figure~\ref{fig:exampletree} written as a set of rules. $I$ is a function denoting the truth of its argument, taking a value of 1 if the conditions apply and a value of 0 if not.}
\label{tab:tree_as_pre_1}
\end{table}

Table~\ref{tab:tree_as_pre_1} also illustrates how a PRE for a continuous response is in fact a linear regression model, where the predictors are rules instead of the original input variables. This is an important characteristic of PREs, as it allows the methodology to be applied to any response variable type that can be modeled within the generalized linear model (GLM). For example, we can employ a logistic regression model for a binary response, or a Poisson regression model for a response reflecting a count of events.

It should be noted that we could have chosen any of the terminal nodes in Figure~\ref{fig:exampletree} as the reference category. In Table~\ref{tab:tree_as_pre_1} we chose the node with a mean close to the overall mean. Instead, we could have chosen the left- or right-most terminal node as the reference category, yielding the parameterizations in panels (a) and (b) of Table~\ref{tab:tree_as_pre_2}, respectively. Taking the left-most terminal node in Figure~\ref{fig:exampletree} as the reference category yields a so-called non-negativity constrained solution, where the intercept reflects the lowest possible predicted value and all rules obtain positive coefficients (Table~\ref{tab:tree_as_pre_2}, panel A). Taking the right-most terminal node as the reference category yields a negativity constrained solution, where the intercept reflects the highest possible predicted value and all rules obtain negative coefficients (Table~\ref{tab:tree_as_pre_2}, panel B). Such alternative parameterizations may be particularly helpful if we want to identify subgroups with higher or lower values of the response variable only. For example, in many applications we may be specifically interested in identifying groups at markedly higher or lower risk for a certain disorder or outcome.

\begin{table}[b]
\centering
\setlength\tabcolsep{2pt}
\begin{tabular}{lc}
\\
\multicolumn{2}{c}{(a) Non-negativity constrained}\\
\\
\hline
Function & Coefficient \\
\hline
Intercept & 2 \\
$I(x_1 \leq 8 \: \& \: x_2 > 10)$ & 3 \\
$I(x_1 > 8)$ & 7 \\
\hline
\end{tabular}
\hfill
\begin{tabular}{lc}
\\
\multicolumn{2}{c}{(b) Negativity constrained}\\
\\
\hline
Function & Coefficient \\
\hline
Intercept & 9 \\
$I(x_1 \leq 8 \: \& \: x_2 \leq 10)$ & -7 \\
$I(x_1 \leq 8 \: \& \: x_2 > 10)$ & -4 \\
\hline
\end{tabular}
\hfill
\caption{Alternative parameterizations for writing the decision tree in Figure~\ref{fig:exampletree} as sets of rules. (a) Parameterization with non-negativity constraint. (b) Parameterization with negativity constraint.}
\label{tab:tree_as_pre_2}
\end{table}

The different parameterizations are similar to the different contrasts for factors that can be used in ANOVAs and regression models, where changing the reference category does not change the conclusions, but may facilitate interpretation. For example, when comparing different treatments in an experimental design, taking the control condition as the reference category simplifies interpretation, as the coefficient for each treatment condition will now reflect the difference with the control condition.

Taking rules instead of the original predictors as inputs provides a number of potential advantages: Firstly, flexibility, as the associations between predictor and response variables need not be linear. Rules can approximate many types of associations, be it monotonically increasing, U-shaped, or interaction effects. Secondly, rules directly identify subgroups of observations with markedly higher or lower values of the response variable. This can be particularly useful in psychological research, where there may be a specific interest in identifying subgroups of persons at high or low risk. Thirdly, rules are easy to interpret and apply. In fact, prediction rules can be seen as fast-and-frugal trees, one of the heuristics postulated by decision scientist Gigerenzer, who found such heuristics closely correspond to the decision-making process of experts, like doctors at emergency units \cite<e.g.,>{FokkySmit15a, GigeyGold96, KatsyPach08, LuanyScho11}.

As noted above, several methods for deriving PREs have been proposed up to date, most exclusively aimed at classification \cite<e.g.,>{CoheySing99, WeisyIndu00}. Such methods may not be widely applicable in psychological research, where continuous, count and other non-categorical response variable types are commonly encountered. PRE methods that allow for the analysis of categorical as well as continuous outcomes are RuleFit \cite{FrieyPope08}, ENDER \cite{DembyKotl10} and Node Harvest \cite{Mein10}.

\subsection{Aim of the Current Paper}

In the current paper we will focus on the RuleFit algorithm of \citeA{FrieyPope08}, as it can be extended to allow for the analysis of a wide range of response variable types and rule generation methods. Also, the RuleFit algorithm has been found to provide predictive performance competitive with state-of-the-art methods like boosted tree ensembles and random forests \cite<e.g.,>{Fokkinpress, FrieyPope08, JolyySchn12, ShimyLi14, YangyZhan08}. Specifically, the current paper will focus on \textbf{R} package \textbf{pre} \cite{Fokkinpress}, which provides a flexible open-source implementation of the RuleFit algorithm and can be applied to a wide range of prediction problems, including the examples mentioned at the beginning of this paper. 

Package \textbf{pre} offers several potential advantages over the original RuleFit implementation \cite{FrieyPope12}: For example, it supports continuous, categorical, count, survival and multivariate responses; it uses unbiased rule induction algorithms; it is well documented and provides a familiar interface for users familiar with \textbf{R}. An extensive description of the original RuleFit algorithm is provided in \citeA{FrieyPope08}. The functionality and implementation of \textbf{pre} has been described in technical detail in \citeA{Fokkinpress}. The current paper aims to provide a less technical introduction and practical tutorial to PREs, aimed at substantive researchers in psychology and related areas.

In the remainder of this paper, we will explain and illustrate PRE methodology through several real-data examples from psychology: In the first example, we will ana\-lyze a dataset on the prediction of chronic depressive disorder among individuals with a current depression. Through this example, we will explain and illustrate how PREs are fitted, and how the results can be interpreted, for example using importance measures and partial dependence plots. In the second example, we will ana\-lyze a dataset on academic achievement among first-year psychology students. Through this example, we will illustrate the analysis of a multivariate response variable and the application of non-negativity constraints. In the third example, we will predict the number of days of substance use in the last week of a clinical trial of substance abuse treatments. Through this example, we will illustrate how a response reflecting a count of events can be modeled, and how rules or variables known a-priori to be predictive of the response (i.e., \textit{confirmatory rules}) can be included in fitting a PRE.

As noted above, the aim of PRE methodology is to strike a balance between the interpretability of single trees and the predictive performance of tree ensembles. Therefore, in the section \textit{Predictive Performance}, we compare the performance of the PREs fitted in the examples with that of single trees, tree ensembles and penalized GLMs. Finally, in the \textit{Discussion}, we will focus on practical issues and questions that may be encountered in fitting PREs, like sample size, computation time and how to optimally choose the model-fitting parameters.    

The examples in the current paper can be replicated using the \textit{Supplementary Material} on \url{https://github.com/marjoleinF/met0000256/}, which provides and explains the \textbf{R} code used to obtain the results presented. In addition, the \textit{Supplementary Material} includes artificially generated datasets, which mimic the datasets used in the examples. Because we do not have permission to redistribute the datasets analyzed in the examples, we have created artificial datasets to allow readers to replicate all analyses presented. 

Due to the introductory nature of the current paper, in the examples we will focus on the default settings employed by package \textbf{pre}. The \textit{Supplementary Material} provides instructions on how to adjust the default model-fitting parameters to optimize complexity and predictive performance of the final ensemble. A more technical and detailed description of possible parameter settings and their effects is provided in \citeA{Fokkinpress}.

\FloatBarrier

\section{Example 1: Predicting Chronic Depression}

\subsection{Dataset}

In this example, we analyze a dataset from the Netherlands Study of Depression and Anxiety \cite<NESDA;>{PennyBeek08}. We use part of a dataset from the analyses performed by \citeA{PennyNole11}, who aimed to predict chronic disorder trajectories among respondents with a current depressive and/or anxiety disorder. \citeA{PennyNole11} operationalized a chronic trajectory as the presence of a disorder at two-year follow-up. They determined predictors of chronicity using logistic regression analyses with 20 potential predictor variables. 

We use the same set of potential predictor variables and analyze a subset of respondents who met the criteria of a depressive disorder at baseline. We omitted 40 respondents with one or more missing values from the dataset, as there are currently no methods to account for missing values in fitting PREs\footnote{Multiple imputation may provide a more appropriate way to deal with missing data \cite<e.g.,>{Grah09}. Every imputed dataset would however yield a (slightly) different PRE, thus yielding a (much) more complex result. In light of the introductory nature of the current paper, we therefore analyzed complete cases only.}. This yielded a total sample size of $N = $ 682. Below, we provide a brief description of the variables; a more detailed explanation of the variables is provided in \citeA{PennyNole11}.

\subsubsection{Psychiatric diagnoses}

All psychiatric diagnoses were assessed according to DSM-IV criteria, by means of a semi-structured diagnostic interview. The response variable is the presence of a depressive disorder at two-year follow-up; 50.59\% of the observations had a depressive disorder at follow-up. Among the potential predictor variables are \textit{disType}, which reflects baseline psychiatric status and distinguishes between two mutually exclusive categories: pure depression, and comorbid depression and anxiety. \textit{TypeDep} reflects the type of depressive disorder at baseline and distinguishes between three mutually exclusive categories: first episode major depressive disorder (MDD), recurrent MDD, and dysthymia. Variables \textit{SocPhob}, \textit{GAD}, \textit{Panic}, and \textit{Ago} reflect the presence of anxiety disorders at baseline: social phobia, generalized anxiety disorder, panic disorder and agoraphobia without panic disorder, respectively. \textit{AO} represents the age of onset of the index disorder.

\subsubsection{Socio-demographic and self-report indicators}

\textit{Sex} (35.48\% male), \textit{Age} (range 18-64), and \textit{edu\_years} (range 5-18) reflect self-reported gender, age in years and years of completed education, respectively. \textit{PsychTreat} and \textit{ADuse} are indicators for whether the respondent was receiving psychological treatment and/or anti-depressant medication at baseline. Variable \textit{pedigree} is an indicator for whether the respondent reported having a first-degree relative with anxiety or depressive disorder. Finally, \textit{sample} is an indicator for whether respondents were originally recruited from primary care, specialized mental health care, or from the general population.

\subsubsection{Symptom severity and duration}

\textit{IDS} is a psychological test score reflecting severity of depressive symptoms. \textit{BAI} and \textit{FQ} are psychological test scores reflecting severity of anxiety symptoms. \textit{LCImax} reflects the proportion of time in which symptoms of anxiety or depressive disorders were present in the four years prior to baseline. Because separate indicators for anxiety and depressive symptoms were collected, \textit{LCImax} reflects the maximum value on both indicators for every respondent.

\subsection{Deriving PREs: Rule Fitting}

To derive an ensemble of prediction rules, the RuleFit algorithm first generates an ensemble of decision trees using subsampling. Package \textbf{pre} takes the same approach: by default, 500 subsamples of 50\% of the original data are drawn. A decision tree is fitted on each of these subsamples. For tree fitting, the original RuleFit algorithm employs the classification and regression tree (CART) algorithm of \citeA{BreiyFrie84}. However, CART presents with biased variable selection: Given two variables equally predictive of the outcome, CART is more likely to select the variable with a larger number of unique values for splitting \cite<e.g.,>{LohyShih97, HothyHorn06}. Therefore, \textbf{pre} by default employs the conditional inference tree (ctree) algorithm of \citeA{HothyHorn06}, which implements an unbiased variable selection procedure. 

To limit rule complexity, \textbf{pre} fits trees with a maximum depth of three, yielding rules comprising at most three conditions, by default. Whereas the original RuleFit algorithm randomly generates a maximum tree depth value for every tree from a pre-specified distribution, this appears not to improve sparsity or predictive performance of the resulting ensemble \cite{Fokkinpress}.

The RuleFit algorithm generates trees sequentially: every tree is grown on a residualized version of the response variable. That is, the response variable is adjusted for the predictions of earlier trees. This procedure is known as (gradient) boosting \cite{FreuyShap95, Frie01}. In boosting, the influence of earlier trees is controlled by the learning rate: a learning rate of 0 yields no influence of earlier trees, while a value of 1 maximizes the influence of earlier trees. \citeA{FrieyPope03} found small, non-zero learning rates to perform best, which is reflected in both RuleFit's and \textbf{pre}'s default learning rate of .01.

\subsection{Deriving PREs: Rule Selection}

After the decision tree ensemble is generated, the RuleFit algorithm includes every node from every tree as a rule in the initial rule ensemble. In addition, \textbf{pre} by default retains only one rule from sets of perfectly positively or negatively correlated rules. This does not affect predictive performance of the final model, but likely reduces complexity and computation time.

As in the original RuleFit algorithm, \textbf{pre} includes all predictor variables as linear terms in the initial ensemble, by default. This is because a large number of rules would be required to approximate a purely linear function of a continuous predictor variables. Thus, if the effect of a continuous predictor can be well approximated by a linear function, the linear term can be selected instead of multiple rules. To reduce the effect of possible outliers, predictor variables are winsorized prior to their inclusion as linear terms in the initial ensemble. That is, all values below the 5th percentile are set to the value of the 5th percentile, and all values above the 95th percentile are set to the value of the 95th percentile, by default. 

The RuleFit algorithms selects the final ensemble through regressing the response variable on all rules and/or linear terms in the initial ensemble. To obtain a sparse final ensemble (i.e., an ensemble in which many rules and linear terms obtain an estimated regression weight of 0), the RuleFit algorithm employs lasso regression. Similarly, package \textbf{pre} employs lasso regression as implemented in the \textbf{R} package \textbf{glmnet} \cite{FrieyHast10, SimoyFrie11}, by default. 

An extensive discussion of lasso regression is outside the scope of the current paper, but an introduction aimed at researchers in psychology can be found in \citeA{ChapyWeis16}. In short, in traditional regression models, coefficients are generally estimated by minimizing the loss function (e.g., residual sum of squares or negative log likelihood). In lasso regression, a penalty term, which is a function of the sum of the estimated coefficients, is added to the loss function. As a result, in order to minimize the loss function, the estimated regression coefficients are shrunken towards zero, compared to the unpenalized estimates. One of the advantages of penalizing the sum of the regression coefficients is that unimportant or noisy predictors may obtain weights shrunken to exactly zero, removing them from the model completely.

The amount of shrinkage applied is controlled by a penalty parameter $\lambda$, which takes values between 0 and 1. A value of 0 yields an unpenalized solution, a value of 1 yields an intercept-only solution. The optimal $\lambda$ value for a given dataset is best determined through cross-validation techniques.

The original RuleFit implementation \cite{FrieyPope12} by default employs the value of $\lambda$ that minimizes prediction error as assessed by $k$-fold cross validation. Furthermore, it determines the number of folds ($k$) based on sample size, with $k$ varying between 1 and 20 and higher sample sizes yielding a smaller number of folds. In contrast, \textbf{pre} by default employs 10-fold cross validation, and the value of $\lambda$ that yields a cross-validated prediction error within one standard error above the minimum cross-validated prediction error. This '1-SE' rule is a heuristic that selects the simplest model with predictive performance comparable to that of the best model (\citeNP{BreiyFrie84}, section 3.4.3). In \citeA{Fokkinpress}, the 1-SE rule based on 10-fold cross validation was found to yield final ensembles that are less complex, while yielding equal or better predictive performance.

Because the lasso penalizes predictors with smaller variance more heavily, both the original RuleFit implementation and \textbf{pre} scale all linear terms to have a standard deviation of 0.4 (i.e., the standard deviation of a typical rule), prior to estimation of the final ensemble.

\subsection{Fitted Ensemble for Predicting Chronic Depression}

The fitted PRE for predicting depression is presented in Table~\ref{tab:terms.NESDA.ens}. Predictive performance, assessed using cross validation and quantified by the area under the receiver operating characteristic curve (AUC) was .641. This result will be discussed in more detail in the section \textit{Predictive Performance}. 

The coefficients reported in Table~\ref{tab:terms.NESDA.ens} are logistic regression coefficients: they represent the expected increase in the log-odds of belonging to the target class (i.e., chronic depression trajectory) for a unit increase in the predictor variable, while keeping all other predictors constant. The value of the intercept thus indicates that observations that do not match the conditions of any of the rules have a probability of $\frac{e^{-0.218}}{1 + e^{-0.218}} = 0.446$ of chronic depression.

\begin{table}[b]
\small
\begin{tabular}{rclcc}
  \hline
Term & Coefficient & Description & SD & Importance \\ 
  \hline
(Intercept) & -0.218 & 1 &  &  \\ 
  rule3 & 0.246 & IDS > 10  \&  LCImax > 0.263 & 0.494 & 0.121 \\ 
  rule27 & 0.152 & IDS > 13  \&  LCImax > 0.362 & 0.477 & 0.073 \\ 
  rule67 & 0.140 & IDS > 10  \&  LCImax > 0.328 & 0.500 & 0.070 \\ 
  rule84 & -0.083 & IDS $\leq$ 16  \&  AO > 17 & 0.489 & 0.041 \\ 
  rule51 & 0.016 & LCImax > 0.260  \&  IDS > 9 & 0.487 & 0.008 \\ 
  rule24 & -0.005 & IDS $\leq$ 16  \&  GAD $\in$ \{Negative\} & 0.499 & 0.002 \\ 
   \hline
\end{tabular}
\caption{Selected terms for predicting chronic depression with estimated coefficients, standard deviations and importances.}
\label{tab:terms.NESDA.ens}
\end{table}

The first rule (\textit{rule3}) indicates that if respondents have an \textit{IDS} value $> 10$ and had symptoms of depression or anxiety for $> 26.3$\% of the time, their odds of having a depression after two years increase. The second and third rules indicate a very similar pattern. The fourth rule (\textit{rule84}) indicates that if respondents have an \textit{IDS} value $\leq 16$ and first met the criteria of the index disorder at age $> 17$, their odds of having a depression in two years time decrease.

Most rules in Table~\ref{tab:terms.NESDA.ens} involve the \textit{IDS} and \textit{LCImax} variables, indicating that these are the most important predictors of having a depressive disorder in two years time. The fitted PRE thus reveals a highly plausible pattern, where more severe depressive symptomatology and longer symptom duration yield a higher predicted probability of a chronic trajectory.

\subsection{Generating Predictions}

To generate predictions for new observations, the contributions of each rule and linear term in the final ensemble need to be computed and summed. An example for two observations is provided in Table \ref{tab:explain.NESDA.ens}. In panel (a), the input variable values are presented. Only the values for those variables appearing in the final ensemble are provided, because only these are required for making a prediction. In panel (b), the coded rules are presented: they take a value of 1 if the conditions of the rule apply and 0 if not. In panel (c), the rules are multiplied by their respective coefficients. To obtain the predicted probability of a chronic depressive trajectory, these values are summed and the logistic function is applied to this sum, yielding the predicted probability of a chronic depressive trajectory. 

Observation 48 has relatively low values of \textit{IDS} and \textit{LCImax}, and met the conditions of both rules with negative coefficients, thus obtaining a lower predicted probability of a chronic depressive trajectory. Observation 64 has relatively high values of \textit{IDS} and \textit{LCImax}, and met the conditions of all rules with positive coefficients, thus obtaining a higher predicted probability of a chronic depressive trajectory.

\begin{table}[h]
\setlength{\tabcolsep}{3pt}
\small
(a) Original variable values\vspace{.2cm}\\
\begin{tabular}{rcccc}
  \hline
 & IDS & LCImax & AO & GAD \\ 
  \hline
48 & 12 & 0.038 & 33 & Negative \\ 
  64 & 20 & 0.722 & 28 & Positive \\ 
   \hline
\end{tabular}\\(b) Rules\vspace{.2cm}\\
\begin{tabular}{rcccccc}
  \hline
 & rule3 & rule24 & rule27 & rule51 & rule67 & rule84 \\ 
  \hline
48 &    0 &    1 &    0 &    0 &    0 &    1 \\ 
  64 &    1 &    0 &    1 &    1 &    1 &    0 \\ 
   \hline
\end{tabular}\\(c) Contributions to the predicted values\vspace{.2cm}\\
\begin{tabular}{rccccccccc}
  \hline
 & (Intercept) & rule3 & rule24 & rule27 & rule51 & rule67 & rule84 & sum & $\hat{p}$ \\ 
  \hline
48 & -0.218 & 0.000 & -0.005 & 0.000 & 0.000 & 0.000 & -0.083 & -0.306 & 0.424 \\ 
  64 & -0.218 & 0.246 & 0.000 & 0.152 & 0.016 & 0.140 & 0.000 & 0.335 & 0.583 \\ 
   \hline
\end{tabular}
\caption{(a) Original variable values, (b) rules, and (c) contributions to the predicted probability of a chronic depressive trajectory for two observations (with identification numbers 48 and 64).}
\label{tab:explain.NESDA.ens}
\end{table}

\subsection{Interpreting PREs: Importance Measures}

The effect of rules and linear terms on the ensemble's predictions are quantified by their estimated coefficients. However, the strength of the effect of a rule or linear term on the ensemble's prediction does not only depend on the estimated coefficient; it also depends on how much the values of a predictor variable or membership of a rule vary over observations. Variables that show more variation in the training data can be expected to have a stronger effect on the ensemble's predictions. \citeA{FrieyPope08} therefore defined several importance measures, which quantify the relative contribution of rules, linear terms and predictor variables to the predictions of the final ensemble. 

\citeA{FrieyPope08} defined the importance of a rule or linear term as the absolute value of the estimated regression coefficient, multiplied by its sample standard deviation. Thus, these importances can take values $\geq 0$, with higher values indicating a stronger effect on the ensemble's predictions. The standard deviation of a rule is given by the standard deviation of a binary variable; that is, $\sqrt{p (1-p)}$, where $p$ is the proportion of training observations to which the rule applies. Thus, the importance of a rule increases both with the absolute value of the estimated coefficient, as well as with the extent to which rule membership varies among the training observations. Package \textbf{pre} computes these importances in the same way, and the rule importances for predicting chronic depressive trajectories are also presented in Table~\ref{tab:terms.NESDA.ens}.

A single predictor variable may appear in the conditions of multiple rules and as a linear term. \citeA{FrieyPope08} therefore defined the total importance of a predictor variable as the sum of the importance of the linear term and the importance of every rule in which the variable appears, divided by the total number of conditions in the rule. Package \textbf{pre} computes these importances in the same way. 

\begin{figure}[h]
\setkeys{Gin}{width=0.5\textwidth}
\centering
\small
\includegraphics{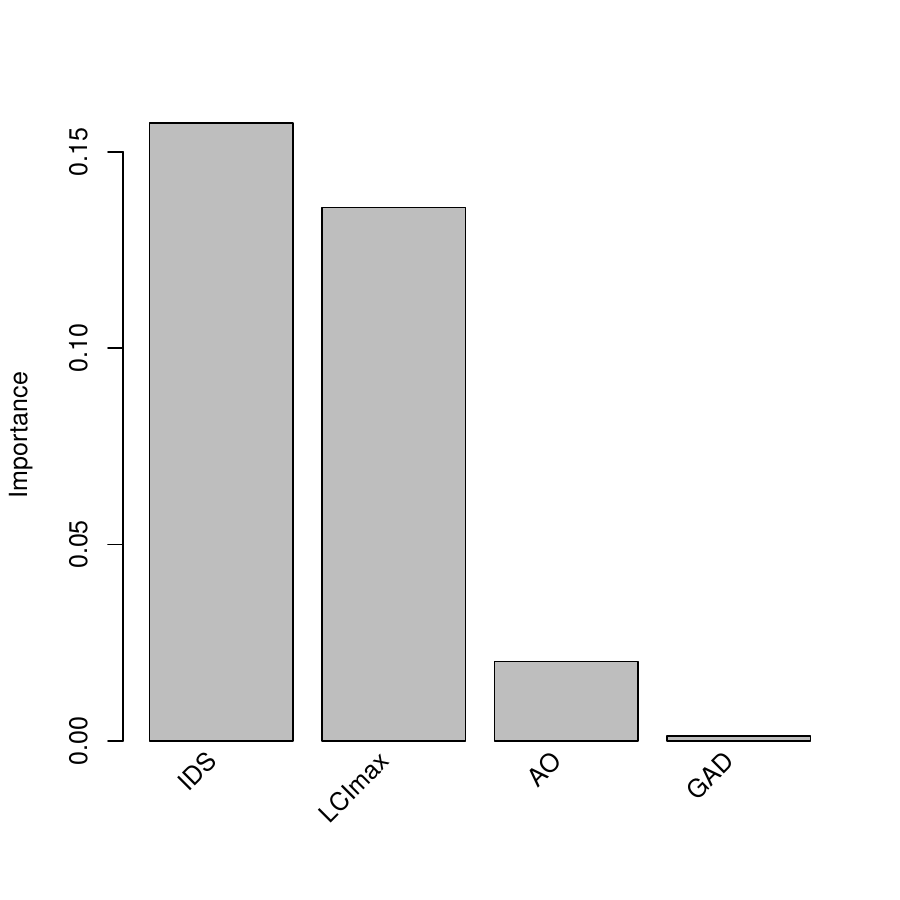}
\caption{Variable importances for predicting chronic depression. Variables with importances equal to 0 (i.e., variables that do not contribute to the predictions of the ensemble) are not plotted.}
\label{fig:imp.NESDA.ens}
\end{figure}

Figure~\ref{fig:imp.NESDA.ens} depicts the individual variable importances for predicting chronic depression. The importances indicate that 4 variables are relevant for predicting depression; the remaining potential predictor variables obtained importance values of 0 and are therefore not included in the plot. As we already observed in Table~\ref{tab:terms.NESDA.ens}, the \textit{IDS} and \textit{LCImax} variables are most important for predicting depression. Age of disorder onset and presence of generalized anxiety disorder are also relevant for predicting chronic depression, but of lesser importance.

\FloatBarrier

\subsection{Interpreting PREs: Partial Dependence Plots}

Partial dependence plots provide visual tools to assess the shape of the effect of one or two predictor variables on the predictions of the ensemble. Partial dependence plots show the conditional expectation of the response variable for a range of values of (pairs of) predictor variables, calculated over the marginal joint distribution of the remaining predictor variables \cite{Frie01, FrieyPope08}. A word of caution is in order here: Because the expectations are calculated over the marginal joint distribution of the remaining predictor variables, possible interactions will be averaged over. A univariate partial dependence plot thus depicts the univariate effect of a predictor, which should be interpreted with care in the presence of interactions. Similarly, a bivariate partial dependence plot can only depict the bivariate effect of a pair of predictors, and should be interpreted with care in the presence of interactions with the remaining predictor variables. The possible presence of (higher-order) interactions in the fitted model can best be evaluated by inspecting rules in the ensemble involving multiple predictor variables.

\begin{figure}[H]
\setkeys{Gin}{width=\textwidth}
\centering
\small
\includegraphics{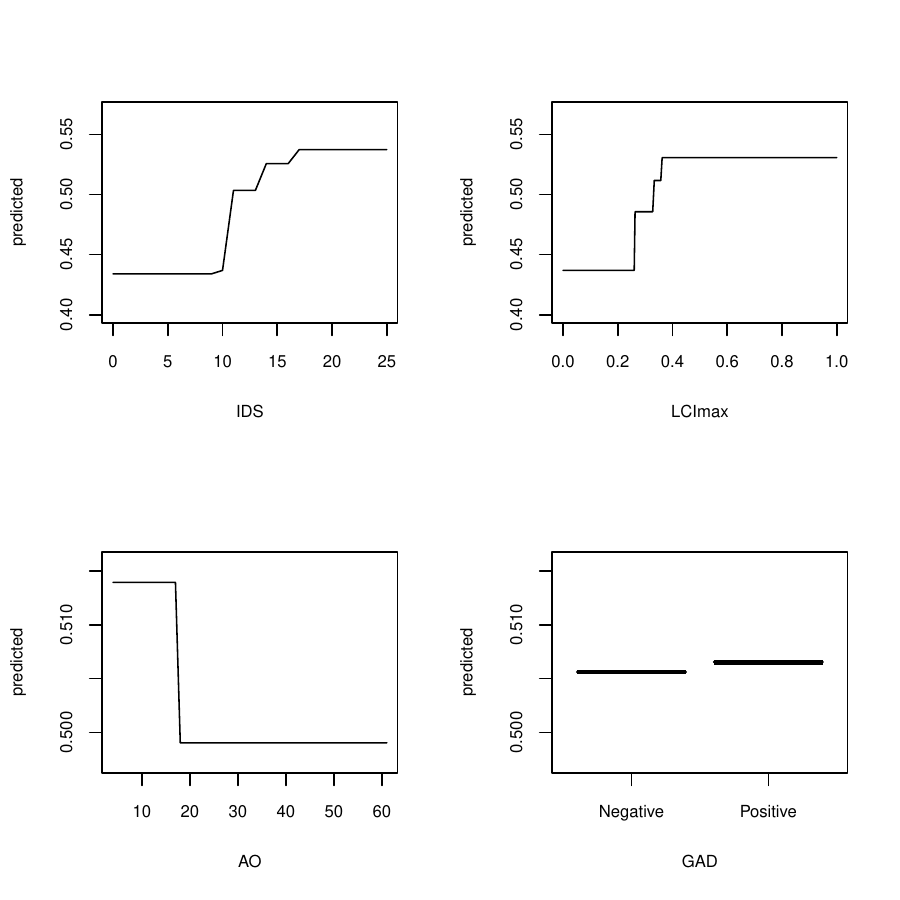}
\caption{Univariate partial dependence of the predicted probability of a depression diagnosis in two years time on four of the predictor variables.}
\label{fig:NESDA.ens.partdep}
\end{figure}

Like the original RuleFit implementation, package \textbf{pre} computes partial dependence plots following the definition of \citeA{FrieyPope08}. Figure~\ref{fig:NESDA.ens.partdep} shows partial dependence plots for a subset of variables relevant for predicting the presence of depressive disorder. The plots reveal monotonically increasing effects of \textit{LCImax} and \textit{IDS} on the predicted probability of chronic depression. Age of disorder onset (\textit{AO}) has a monotonically decreasing effect on the probability of having chronic depression, while the presence of a generalized anxiety disorder slightly increases the probability.

\begin{figure}[H]
\setkeys{Gin}{width=0.7\textwidth}
\centering
\small
\includegraphics{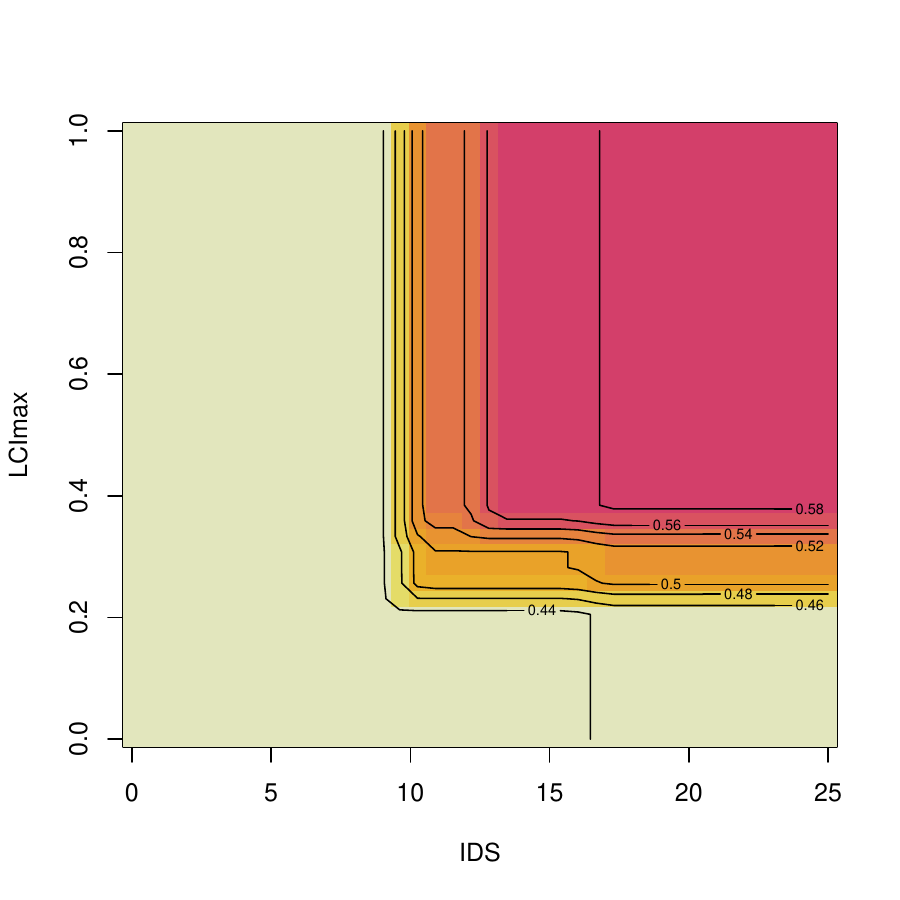}
\caption{Bivariate partial dependence of the predicted probability of chronic depression on depressive symptom severity (\textit{IDS}) and symptom duration (\textit{LCImax}).}
\label{fig:NESDA.ens.pairplot}
\end{figure}

Figure~\ref{fig:NESDA.ens.pairplot} shows the dependence of the probability of having a chronic depression on both symptom severity (\textit{IDS}) and duration (\textit{LCImax}). The plot reveals a pattern where the probability of a chronic trajectory increases with both increasing symptom severity and symptom duration.

\newpage 

\FloatBarrier

\section{Example 2: Predicting Academic Achievement}

In this second example, we focus on the prediction of a multivariate response and on the application of non-negativity constraints. We will be using a dataset on academic achievement, collected from first-year bachelor students in psychology at Groningen University (The Netherlands).

\subsection{Dataset}

The data are from a study by \citeA{NiesyMeij16}, where the variables are described in more detail. Below, we provide a brief description of the variables used in our analyses. The original dataset comprised 851 observations, after removing observations with one or more missing values, we obtained a sample of $N = $ 638 observations. We apply a non-negativity constraint, which will force the estimated coefficients of rules and linear terms in the final PRE to values $\geq 0$. Thus, the rules in the final ensemble will only identify subgroups of students who are likely to perform better than other students. Note that the application of (non-)negativity constraints is not limited to multivariate or continuous responses, but can be applied with any response variable type.

\subsubsection{Response variables} 

The response consists of two variables: \textit{MeanFYG}, representing the mean grade over all courses in the first year (range 1.0-9.4) and \textit{Credits}, representing the number of credits obtained in the first year (range 0-60). The sample correlation between these two response variables was 0.82.

\subsubsection{Socio-demographic variables} 

Potential socio-demographic predictor variables are \textit{Gender} (30.41\% male), \textit{Age} (range 17-42), \textit{Nationality} (Dutch, German, other EU country, or Non-EU country). \textit{Program} represents an indicator for whether students participated in the Dutch or English study program.

\subsubsection{Admission tests} 

Before the start of their studies, prospective students completed three admission tests, yielding three test scores: \textit{RawScore\_English}, \textit{RawScore\_Math}, and \textit{RawScore\_Psychology}. \textit{Online\_test} represents an indicator for whether admission tests were completed online or in person; \textit{Test\_Language} represents an indicator for whether admission tests were taken in English or in Dutch.

\subsection{Fitted Ensemble for Predicting Academic Achievement}

\begin{table}[h]
\setlength\tabcolsep{2pt}
\small
\begin{tabular}{rrcp{9cm}}
  \hline
Term & Credits & MeanFYG & Description \\ 
  \hline
(Intercept) & 30.425 & 5.384 & 1 \\ 
  rule21 & 5.124 & 0.317 & RawScore\_Psychology > 24  \&  Nationality $\in$ \{Dutch, German\} \\ 
  rule206 & 3.126 & 0.231 & RawScore\_Psychology > 24  \&  RawScore\_Math > 12 \\ 
  rule442 & 2.113 & 0.108 & Nationality $\in$ \{Dutch, German\}  \&  Gender $\in$ \{female\} \\ 
  rule316 & 1.371 & 0.107 & RawScore\_Psychology > 24  \&  RawScore\_Math > 13 \\ 
  rule40 & 1.162 & 0.100 & RawScore\_Psychology > 30 \\ 
  rule352 & 1.311 & 0.088 & RawScore\_Psychology > 24  \&  RawScore\_Math > 10 \\ 
  rule444 & 1.569 & 0.086 & Nationality $\in$ \{Dutch, German\}  \&  Gender $\in$ \{female\}  \&  RawScore\_Psychology > 22 \\ 
  rule12 & 0.758 & 0.083 & RawScore\_Psychology > 24  \&  Nationality $\in$ \{German\} \\ 
  rule150 & 0.645 & 0.059 & RawScore\_Psychology > 24  \&  RawScore\_Psychology > 31 \\ 
  rule618 & 0.308 & 0.032 & RawScore\_Math > 15  \&  RawScore\_Psychology > 28 \\ 
  RawScore\_Psychology & 0.176 & 0.016 & 18 $\leq$ RawScore\_Psychology $\leq$ 38 \\ 
  rule258 & 0.133 & 0.011 & RawScore\_Psychology > 18  \&  RawScore\_Math > 12  \&  RawScore\_Psychology > 28 \\ 
   \hline
\end{tabular}
\caption{Selected terms and estimated coefficients for predicting mean first-year grade and the number of credits obtained.}
\label{tab:terms.nies.ens.mv}
\end{table}

The fitted ensemble is presented in Table~\ref{tab:terms.nies.ens.mv}. Note that, with a multivariate response, every rule (or linear term) obtains an estimated coefficient for each response variable. Because we applied a non-negativity constraint, all estimated coefficients are positive. As the two response variables are strongly correlated, the estimated coefficients have a similar order of magnitude for the two response variables. The ensemble's coefficients of determination ($R^2$) for predicting the number of credits was 0.221, and 0.272 for predicting the mean first-year grade, as assessed by cross validation. In the section \textit{Predictive Performance}, this result will be discussed in more detail.

The Psychology admission test score appears to be an important predictor for academic achievement: It appears in the conditions of almost every rule of the final ensemble and also as a linear term. Note that the description for the linear term (i.e., \textit{RawScore\_Psychology} in Table~\ref{tab:terms.nies.ens.mv}) also provides the winsorizing points for this variable. That is, values of \textit{RawScore\_Psychology} $< 18$ were set to 18 and values $> 38$ were set to 38, to reduce the effect of possible outliers. The terms involving \textit{RawScore\_Psychology} reveal a highly plausible pattern, where higher values on the admission test are associated with higher academic achievement.

Furthermore, the score on the Math admission test, nationality and gender also contribute to the ensemble's predictions. A possible explanation for the effect of nationality is that Dutch and German students are more accustomed to the Dutch language and academic environment, compared to students from other (EU and non-EU) countries.

\subsection{Generating Predictions}

Input variable values for two observations are presented in panel (a) of Table \ref{tab:explain.niessen.ens}. In panel (b), the rules are coded. In panel (c), the rules and linear term are multiplied by their respective coefficients, and summed to obtain a prediction for each response variable.

\begin{table}
\setlength{\tabcolsep}{2pt}
\small
\vspace{.6cm}
(a) Original variable values\vspace{.2cm}\\
\begin{tabular}{rcccc}
  \hline
 & RawScore\_Psychology & Nationality & RawScore\_Math & Gender \\ 
  \hline
31 & 16 & German & 8 & male \\ 
  387 & 38 & German & 16 & female \\ 
   \hline
\end{tabular}\\(b) Rules\vspace{.2cm}\\
\begin{tabular}{rccccccccccc}
  \hline
 & rule12 & rule21 & rule40 & rule150 & rule206 & rule258 & rule316 & rule352 & rule442 & rule444 & rule618 \\ 
  \hline
31 &    0 &    0 &    0 &    0 &    0 &    0 &    0 &    0 &    0 &    0 &    0 \\ 
  387 &    1 &    1 &    1 &    1 &    1 &    1 &    1 &    1 &    1 &    1 &    1 \\ 
   \hline
\end{tabular}\\(c) Contributions to the predicted values of \textit{Credits} and \textit{MeanFYG}\vspace{.2cm}\\
\begingroup\scriptsize
\begin{tabular}{rcccccccccccccc}
  \hline
 & (Intcpt) & RSc\_Psy & rule12 & rule21 & rule40 & rule150 & rule206 & rule258 & rule316 & rule352 & rule442 & rule444 & rule618 & $\hat{y}$ \\ 
  \hline
31 & 30.425 & 3.162 & 0.000 & 0.000 & 0.000 & 0.000 & 0.000 & 0.000 & 0.000 & 0.000 & 0.000 & 0.000 & 0.000 & 33.587 \\ 
  387 & 30.425 & 6.674 & 0.758 & 5.124 & 1.162 & 0.645 & 3.126 & 0.133 & 1.371 & 1.311 & 2.113 & 1.569 & 0.308 & 54.720 \\ 
   \hline
\end{tabular}
\endgroup
\begingroup\scriptsize
\begin{tabular}{rcccccccccccccc}
  \hline
 & (Intcpt) & RSc\_Psy & rule12 & rule21 & rule40 & rule150 & rule206 & rule258 & rule316 & rule352 & rule442 & rule444 & rule618 & $\hat{y}$ \\ 
  \hline
31 & 5.384 & 0.290 & 0.000 & 0.000 & 0.000 & 0.000 & 0.000 & 0.000 & 0.000 & 0.000 & 0.000 & 0.000 & 0.000 & 5.674 \\ 
  387 & 5.384 & 0.611 & 0.083 & 0.317 & 0.100 & 0.059 & 0.231 & 0.011 & 0.107 & 0.088 & 0.108 & 0.086 & 0.032 & 7.218 \\ 
   \hline
\end{tabular}
\endgroup
\caption{(a) Original variable values, (b) rules, and (c) their contributions to the predicted number of credits and mean grade in the first year for two observations (with identification numbers 31 and 387).}
\label{tab:explain.niessen.ens}
\end{table}

Observation 31, who is male and has relatively low values for \textit{RawScore\_Psychology} and \textit{RawScore\_Math}, obtained a lower predicted number of credits and mean grade for the first year. Observation 387, who is female and has relatively high values for \textit{RawScore\_Psychology} and \textit{RawScore\_Math}, obtained a higher predicted number of credits and mean grade for the first year.

\FloatBarrier

\subsection{Interpreting PREs: Standardized Importance Measures}

To further interpret the fitted ensemble, we can inspect variable importances. Earlier, we explained how importances of rules and linear terms are given by the estimated coefficients, multiplied (i.e., standardized) by the sample standard deviations. To obtain a measure of effect size that can be compared between different response variables, package \textbf{pre} allows for further standardizing the importances, by dividing them by the standard deviation of the response variable. These standardized importances can be interpreted as standardized regression coefficients, taking values between 0 (no correlation), and 1 (perfect correlation), which may further simplify interpretation. These standardized importances are particularly useful when there are multiple response variables with different scales, as in the current example.

\begin{figure}[h]
\setkeys{Gin}{width=0.5\textwidth}
\centering
\small
\includegraphics{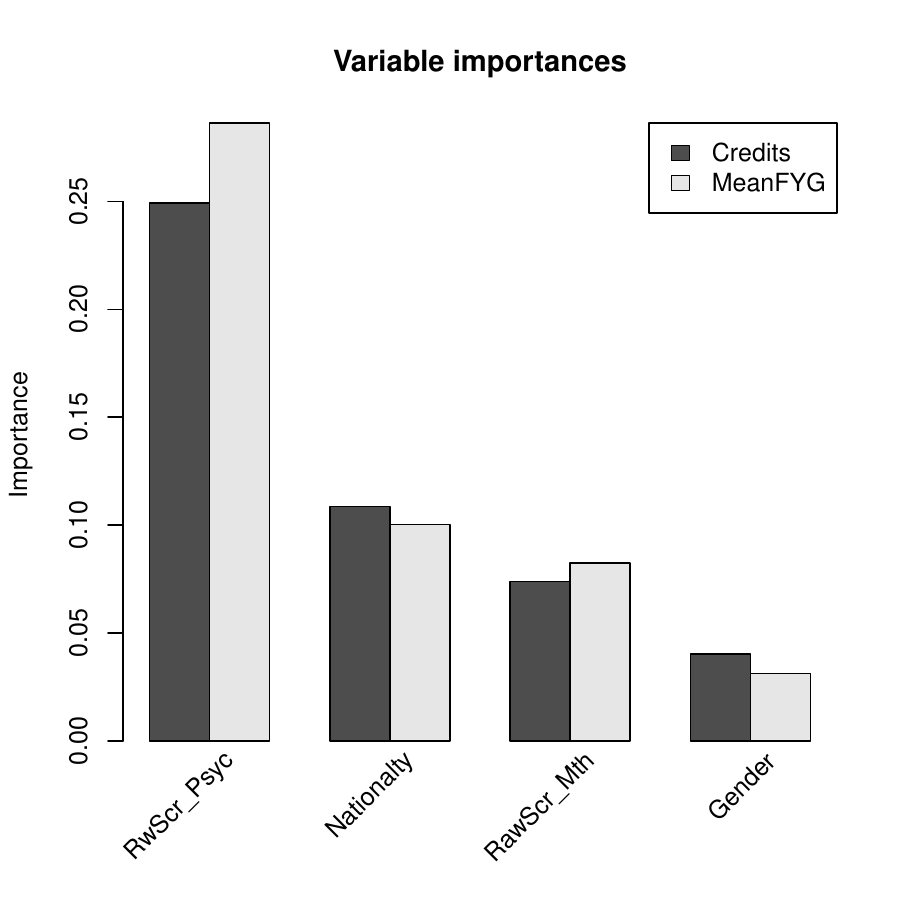}
\caption{Standardized variable importances for predicting mean grade and number of credits obtained in the first year.}
\label{fig:importance.nies.ens.mv}
\end{figure}

In turn, the standardized importances of rules and linear terms can again be used to compute standardized input variable importances. The standardized input variable importances for predicting academic achievement are depicted in Figure~\ref{fig:importance.nies.ens.mv}. The importances reveal a similar pattern for both response variables: The most important predictor variable is the Psychology admission test score, followed by nationality, the Math admission test score, and gender.

\FloatBarrier

\section{Example 3: Predicting Substance Use}

In the third example, we focus on the prediction of a count response and on the inclusion of a confirmatory rule. That is, a rule which is known a-priori to be predictive of the response, for example from previous studies. By specifying a confirmatory rule in fitting a PRE, no penalty will be applied to the estimated coefficient of this rule, and the rule will thus receive an unbiased regression coefficient estimate. 

\subsection{Dataset}

We analyze a dataset from a multi-site effectiveness trial on substance abuse treatment from \citeA{CampyNune12}. The trial was conducted within the National Institute on Drug Abuse (NIDA) Clinical Trials Network (CTN) and the data were obtained from NIDA's data share website on \url{https://datashare.nida.nih.gov/}. The study was a randomized trial, comparing the use of a so-called Therapeutic Education System (TES) with Treatment as Usual (TAU). TAU comprised standard treatment in an outpatient substance abuse program, TES additionally comprised a computer-assisted intervention and contingency management system.

In our analyses, we take the number of reported days of substance use (drugs or alcohol) in the last week of treatment as the response variable. The number of days of substance use reflects a count of events and we therefore model the response through a Poisson regression. 

\citeA{CampyNune14} found that TES resulted in higher abstinence rates than TAU; we therefore expect an effect of receiving TES and include it as a confirmatory rule in fitting the ensemble. As a result, no penalty will be applied to the estimated coefficient of the treatment indicator (\textit{trt}), yielding an unbiased estimate of the effect of TES.

We included a total of 56 potential predictor variables in our analysis, which are briefly described below. A total of 507 patients were randomized to treatment, originally. We included observations with complete data only, yielding a total sample size of $N =$ 478.

\subsubsection{Socio-demographic variables} 
These included patients' age in years (range 18-68), gender (61.51\% male), level of education, marital status, employment status, and living arrangement.

\subsubsection{Quality of life} 
Six self-reported quality of life indicators were included as potential predictors: an overall quality of life rating (on a scale of 0-100), mobility problems, problems with self-care, problems with daily activities, pain or discomfort, and feelings of anxiety or depression. 

\subsubsection{Coping strategies}
Coping strategies in dealing with drugs and alcohol were measured with the 23-item Coping Strategies Scale (CSS; \citeNP{LittyKadd03}). All 23 item responses were included as individual potential predictor variables. 

\subsubsection{Mental health problems}
Mental health problems were measured by the 18-item Brief Symptom Inventory (BSI; \citeNP{DeroyFitz04}). All 18 item responses were included as individual potential predictor variables.

\subsubsection{Clinical characteristics}
These included the type of substance primarily abused by the patient, an indicator for treatment condition, and the number of days the patient used drugs or alcohol in the first week of treatment.

\subsection{Fitted Ensemble for Predicting Substance Use}

\begin{table}[b]
\small
\begin{tabular}{rcp{8cm}}
  \hline
Term & Coefficient & Description \\ 
  \hline
(Intercept) &  0.559 & 1 \\ 
  rule20 & -0.195 & week1 $\leq$ 0  \&  BSNAUSE.T0 $\leq$ 2 \\ 
  trt $\in \{$TES$\}$ & -0.177 & trt $\in \{$TES$\}$ \\ 
  rule16 & -0.157 & week1 $\leq$ 2  \&  CSCALM.T0 > 2 \\ 
  rule30 & -0.120 & week1 $\leq$ 0  \&  BSTENSE.T0 $\leq$ 3 \\ 
  week1 &  0.060 & 0 $\leq$ week1 $\leq$ 7 \\ 
   \hline
\end{tabular}
\caption{Selected terms and estimated coefficients for predicting days of substance use in last week of treatment.}
\label{tab:terms.NIDA.ens}
\end{table}

We obtained a final ensemble consisting of five rules, which are presented in Table~\ref{tab:terms.NIDA.ens}. Note that the confirmatory rule (\textit{trt $\in \{$TES$\}$}) did not receive a rule number, only exploratory rules do. The (unpenalized) coefficient for the confirmatory rule is negative, indicating that the new treatment appears effective: TES is associated with lower substance use than TAU, as was expected. 

All other terms in the ensemble involve variable \textit{week1}, which represents the number of days the patient used substances in the first week of treatment. The \textit{week1} linear term (Table~\ref{tab:terms.NIDA.ens}) has a positive coefficient, indicating a highly plausible positive association between substance use in the first and last week of treatment. The remaining rules indicate a similar effect for the \textit{week1} variable. Furthermore, \textit{BSNAUSE.T0} is an item of the BSI measuring the extent to which the respondent experiences nausea; \textit{rule20} indicates that lower values on this item, combined with no substance use in the first week of treatment, are associated with lower substance use. \textit{CSCALM.T0} is an item of the CCS reflecting the extent to which the respondent reports trying to calm down when feeling angry; \textit{rule16} indicates that higher values on this item, combined with lower substance use in the first week of treatment, are associated with lower substance use in the last week of treatment. \textit{BSTENSE.T0} is an item of the BSI reflecting the extent to which the respondent reports feeling tense or keyed up; \textit{rule30} indicates that lower values on this item, combined with no substance use in the first week of treatment, are associated with lower substance use in the last week of treatment.

To interpret the coefficients in a Poisson regression, we have to exponentiate. For example, the intercept reveals that patients who do not meet the conditions of any rule are predicted to use substances $e^{0.559} = 1.749$ days in the last week of treatment. Similarly, patients in the TES condition, that do not meet the conditions of any of the other rules in the ensemble, are expected to use substances $e^{0.559 -0.177} = 1.466$ days in the last week of treatment.

The $R^2$ value of the ensemble's predictions, calculated using cross validation, was 0.097. This result will be discussed in more detail in the section \textit{Predictive Performance}.

\subsection{Generating Predictions}

Input variable values for two observations are presented in panel (a) of Table \ref{tab:explain.NIDA.ens}. In panel (b), the rules are coded, and in panel (c), the rules and linear term are multiplied by their respective coefficients. These values are summed and exponentiated to obtain a predicted value for the number of substance use days in the last week of treatment. Observation 5 obtained a low predicted value, because of receiving TES, and reporting no use of substances in the first week of treatment, low feelings of nausea and tenseness, and staying calm when feeling angry. Observation 20 obtained a higher predicted value, because of not receiving TES and reporting seven days of substance use in the first week of treatment.

\begin{table}[H]
\setlength{\tabcolsep}{3pt}
\small
(a) Original variable values\vspace{.2cm}\\
\begin{tabular}{rccccc}
  \hline
 & week1 & trt & BSNAUSE.T0 & CSCALM.T0 & BSTENSE.T0 \\ 
  \hline
5 &    0 & TES &    0 &    3 &    1 \\ 
  20 &    7 & TAU &    3 &    1 &    0 \\ 
   \hline
\end{tabular}\\(b) Rules\vspace{.2cm}\\
\begin{tabular}{rcccc}
  \hline
 & trt $\in \{$TES$\}$ & rule16 & rule20 & rule30 \\ 
  \hline
5 &    1 &    1 &    1 &    1 \\ 
  20 &    0 &    0 &    0 &    0 \\ 
   \hline
\end{tabular}\\(c) Contributions to the predicted values\vspace{.2cm}\\
\begin{tabular}{rcccccccc}
  \hline
 & (Intercept) & week1 & rule16 & rule20 & rule30 & trt $\in \{$TES$\}$ & sum & $\hat{y}$ \\ 
  \hline
5 & 0.559 & 0.000 & -0.157 & -0.195 & -0.120 & -0.177 & -0.089 & 0.915 \\ 
  20 & 0.559 & 0.422 & 0.000 & 0.000 & 0.000 & 0.000 & 0.981 & 2.667 \\ 
   \hline
\end{tabular}
\caption{(a) Original variable values, (b) rules, and (c) their contributions to the predicted number of days of substance use in the last week of treatment for two observations (with identification numbers 5 and 20).}
\label{tab:explain.NIDA.ens}
\end{table}

\FloatBarrier

\subsection{Variable Importances}

\begin{figure}[H]
\setkeys{Gin}{width=0.5\textwidth}
\centering
\small
\includegraphics{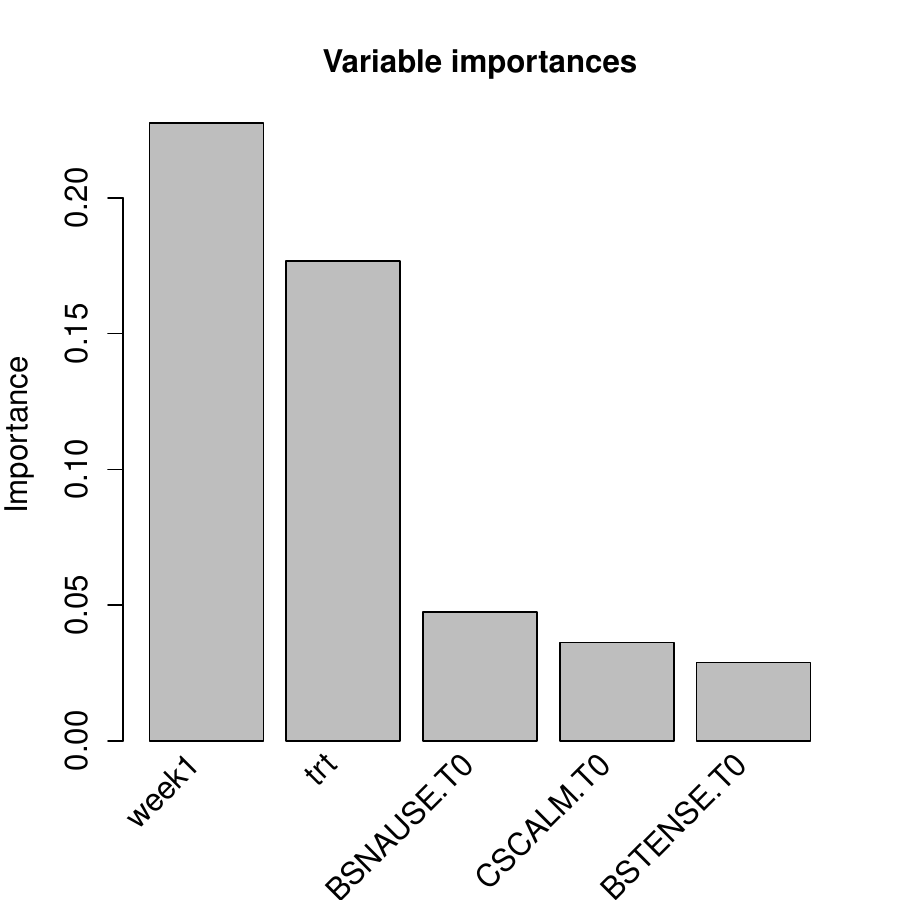}
\caption{Variable importances for predicting substance use days in the last week of treatment.}
\label{fig:importance.NIDA.ens}
\end{figure}

The input variable importances are plotted in Figure~\ref{fig:importance.NIDA.ens}. First-week substance use appears to be a somewhat stronger predictor of last-week substance use than the treatment indicator, even though the latter obtained an unpenalized coefficient. Also note that variable importances quantify the contribution of individual variables to the predictions of the ensemble. Thus, although the estimated coefficient of the treatment indicator was not shrunken towards zero, the estimated coefficients and importances still provide an accurate description of the fitted model and can be compared between predictor variables. Finally, the variable importances in Figure~\ref{fig:importance.NIDA.ens} indicate that the BSI and CSS items have relatively small effects on the predictions of last-week substance use.

\FloatBarrier

\section{Predictive Performance}

As noted in the \textit{Introduction}, PREs aim to strike a balance between the predictive performance of full decision tree ensembles and the interpretability of single decision trees. The PREs fitted in the examples above are indeed less complex than random forests, which generally consist of a large number of decision trees (e.g., 500) and tend to include all predictor variables in the predictive model. Furthermore, we can expect that the fitted PREs are somewhat more complex than single decision trees fitted on the same data would be. In this section, we will compare the predictive performance of the PREs fitted in the examples with that of random forests, single trees and lasso-penalized GLMs.

\subsection{Method}

All analyses were performed in \textbf{R} (\citeNP{R19}, version 3.6.1).

\subsubsection{Random forests}

Random forests were fitted using function \verb|cforest| from package \textbf{partykit} (\citeNP{HothyZeil15}; version 1.2-5). Default settings were employed in all analyses. That is, every random forest consisted of 500 trees, the number of randomly selected predictor variables for each split was set to $\sqrt{p}$ (where $p$ is the number of potential predictor variables), each tree was grown on a random subsample of a fraction of .632 of the original training observations, and splitting was continued as long as a node contained $\geq 20$ observations and both resulting nodes contained $\geq 7$ observations.

\subsubsection{Prediction rule ensembles}

PREs were fitted using function \verb|pre| from package \textbf{pre} (version 0.7.2). Default settings were employed. That is, no (non-)negativity constraints and no confirmatory rules were applied in fitting PREs, as these can also not be applied in fitting single decision trees and random forests.

\subsubsection{GLM with lasso penalty}

Logistic and linear regression models (GLMs) with a lasso penalty were also fitted using function \verb|pre|. Although function \verb|pre| fits ensembles that may include both rules and linear terms, it also allows for fitting ensembles composed of either linear terms or rules. Except for the setting of linear terms only, default settings were employed for fitting the GLMs. That is, outliers in the predictor variables were winsorized to their 5th and 95th percentile, and estimation was performed using function \verb|cv.glmnet| from package \textbf{glmnet} (\citeNP{FrieyHast10}; version 3.0-1) with the '1-SE' rule used for model selection.

\subsubsection{Single decision trees}

Single decision trees were fitted using function \verb|ctree| \cite{HothyHorn06} from package \textbf{partykit}. Default settings were employed in all analyses. That is, an $\alpha$ level of .05 was employed as the significance level for the variable selection test, which determines whether a split will be made or not. Also, splits were made as long as a node contained $\geq 20$ observations and both resulting nodes contained $\geq 7$ observations. 

\subsubsection{Assessment of predictive performance}

Calculating predictive performance based on the same data that was used to construct the model (i.e., training data) likely yields overly optimistic estimates of predictive performance on future observations \cite<e.g.,>{HastyTibs09}. Therefore, predictive performance was calculated based on 10-fold cross validation. That is, the original training observations were randomly assigned to one of 10 (approximately equally-sized) folds. Each of these folds serves as a test dataset, for which predictions are generated using a model build on the observations in the remaining nine folds. This procedure is repeated for each of the 10 folds, yielding a cross-validated prediction of the response variable for each of the original training observations. These cross-validated predictions are then compared with the observed values of the response to assess predictive performance. To reduce the dependence of the results on a single partition of the training observations into folds, we repeated the 10-fold cross validation 10 times. This approach is known as repeated cross validation and has been found to provide more reliable estimates of predictive performance \cite{Kim09, KrstyButu14}.

For the dataset from Example 1 (chronic depression), the squared error loss (SEL) was taken as the outcome. The SEL reflects the mean squared difference between the predicted probability of belonging to the target class and observed class membership, which is sometimes also referred to as the Brier score. For the datasets from Example 2 (multivariate academic performance) and Example 3 (substance use), mean squared error (MSE) was taken as the outcome. The MSE reflects the mean squared difference between the predicted and the observed value of the response variable(s). SEL and MSE were calculated for each fold of each repetition, allowing for computing both the mean predictive performance and its standard error (where $SE = SD / \sqrt{100}$, where 100 equals the number of folds times the number of repetitions).

In addition, measures of effect size were calculated: the area under the receiver operating characteristic curve (AUC) for the depression example, and the coefficient of determination ($R^2$) for the other two examples. For these measures, averages were calculated over the 10 repeats of 10-fold cross validation.

To test the statistical significance of differences in SEL or MSE between the four methods, linear mixed-effects regression models were fitted using package \textbf{lme4} (\citeNP{BateyMach15}; version 1.1-21). The estimates of predictive performance in each of the folds of the repeated 10-fold cross validation represent a repeated-measures experimental design. To account for the dependence introduced by this repeated-measures design, a random intercept was estimated with respect to every fold in every replication. Fixed effects were estimated to compare the performance of PREs with that of random forests and single trees, respectively. Significance of the fixed effects was assessed using package \textbf{lmerTest}, which allows for calculating $p$ values for linear mixed-effect models using Satterthwaite's degrees of freedom method (\citeNP{KutznyBroc17}; version 3.1-0).  

\subsection{Results}

\begin{figure}[h]
\setkeys{Gin}{width=\textwidth}
\centering
\small
\includegraphics{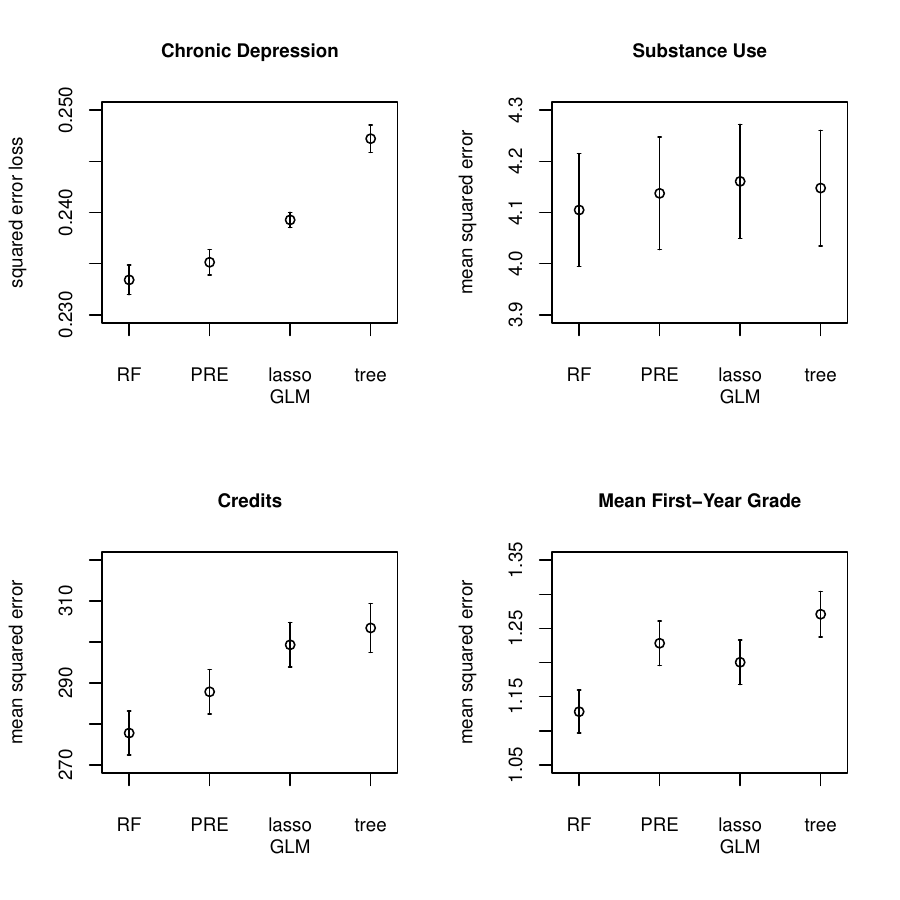}
\caption{Predictive performance for random forests, PREs, lasso-penalized GLMs and single decision trees for the datasets from Example 1 (Chronic Depression), Example 2 (Credits and Mean First-Year Grade) and Example 3 (Substance Use). Circles depict the means and the error bars depict $\pm$ 1 standard error, calculated over 10 replications of 10-fold cross validation. Note that these standard errors do not account for the repeated-measures design of the experiment (as opposed to the multilevel analyses reported in text). RF = random forest, PRE = prediction rule ensemble; lasso GLM = lasso-penalized generalized linear model.}
\label{fig:predictive_accuracy}
\end{figure}

\begin{table}[b]
\small
\centering
\begin{tabular}{lcccc}
  \hline
 & RF & PRE & lasso
GLM & tree \\ 
  \hline
Chronic Depression (AUC) & 0.651 & 0.641 & 0.637 & 0.582 \\ 
  Credits ($R^2$) & 0.249 & 0.221 & 0.190 & 0.179 \\ 
  Mean First-Year Grade ($R^2$) & 0.332 & 0.272 & 0.289 & 0.247 \\ 
  Substance Use ($R^2$) & 0.104 & 0.097 & 0.092 & 0.095 \\ 
   \hline
\end{tabular}
\caption{Mean effect sizes for predictive accuracies of random forests, PREs and single decision trees for the datasets from Example 1 (Chronic Depression), Example 2 (Credits and Mean First-Year Grade) and Example 3 (Substance Use). Means are calculated over 10 replications of 10-fold cross validation. RF = random forest; PRE = prediction rule ensemble; lasso GLM = lasso-penalized generalized linear model.}
\label{tab:accuracies}
\end{table}

Figure~\ref{fig:predictive_accuracy} presents means and standard errors of the prediction error measures. In all datasets, random forests rank highest in terms of predictive performance. Overall, PREs rank highest after that, followed by the lasso GLMs, followed by the single trees. 

In the dataset from Example 1 (chronic depression), prediction error of PREs was significantly lower than that of single trees ($p$ < .001) and the lasso GLMs ($p$ < .001). Prediction error of PREs was higher than that of random forests, but not significantly ($p =$ 0.083). The AUC values (Table~\ref{tab:accuracies}) indicate the same ordering of performance.

In the dataset from Example 2 (academic performance), prediction error of PREs was significantly lower than that of single trees for both Credits ($p$ < .001) and MeanFYG ($p$ < .001). Prediction error of PREs was also significantly higher than that of random forests for both Credits ($p$ < .001) and MeanFYG ($p$ < .001). Prediction error of PREs was significantly lower than that of the lasso GLMs for Credits ($p$ < .001), but significantly higher for MeanFYG ($p =$ 0.008). The proportions of variance explained (Table~\ref{tab:accuracies}) indicate a medium to large effect for the performance of PREs. 

In the dataset from Example 3 (substance use), prediction error of PREs was lower than that of single trees and the lasso GLMs, but not significantly ($p = $ 0.766, $p = $ 0.511, respectively). Prediction error of PREs was higher than that of random forests, but not significantly ($p = $ 0.362). The proportions of variance explained (Table~\ref{tab:accuracies}) indicate a medium effect for the performance of PREs.

In summary, the results indicate that in the examples presented here, PREs indeed strike a balance between the predictive performance of random forests and single decision trees, while at the same time performing better than lasso GLMs, on average. 

\FloatBarrier

\section{Discussion}

\subsection{Summary}

We introduced PRE methodology as a flexible regression method for psychological research. Using \textbf{R} package \textbf{pre}, we fitted PREs for predicting chronic depression, academic performance and substance use. These examples illustrated how rules in a PRE are derived and selected, how predictor variables and different types of response variables should be specified, how (non-)negativity constraints can be applied, and how rules that are known a-priori to be predictive of the response can be included in the ensemble. The examples also showed how the resulting PREs can be interpreted, for example using importance measures and partial dependence plots. 

We compared predictive performance of the fitted PREs with that of random forests and single decision trees using cross-validation techniques. We found that PREs indeed strike a balance between tree ensembles and single decision trees. This is in line with earlier results on the RuleFit algorithm \cite{FrieyPope08, JolyySchn12, ShimyLi14, YangyZhan08}. We also found that PREs outperformed lasso-penalized GLMS in three out of the four response variables, which is in line with the findings of \citeA{Fokkinpress}, where \textbf{pre} provided lower complexity and higher predictive performance than both lasso-penalized GLMs and the original RuleFit implementation. 

We conclude that PREs may provide flexible, accurate and interpretable tools for prediction problems in psychological research. PREs may be particularly useful for researchers interested in subgroup detection, for example to decide which persons are at high (or low) risk for adverse (or beneficial) outcomes. Whereas traditional GLMs identify which variables are predictive of the outcome, for generating predictions they require the multiplication of predictor variables by their respective coefficients and summing the resulting values. With an increasing number of predictor variables, this requires increasing amounts of information, time and computation, which are generally scarce and costly in psychological practice. Compared to traditional GLMs, the variable selection and subgroup identification performed by PREs may thus be very practically useful. The rules in a PRE can be seen as fast-and-frugal decision trees and may provide easy-to-use heuristical tools for human decision makers \cite<e.g.,>{KatsyPach08, GigeyGold96, LuanyScho11, FokkySmit15a}. 

In the remainder of this section, we will discuss a number of practical issues and questions that may be encountered in fitting PREs:

\subsection{Sample Size}

There are no minimum sample size requirements for fitting PREs; because PREs are an exploratory regression method, there is no concept of statistical power. However, as with all statistical methods, a trade-off between sample size and predictive performance applies, in that larger samples will yield more accurate, generalizable, and stable models. Our results indicate that with sample sizes ranging from about 450 to 650, predictive performance of PREs already approximates that of random forests, which are a state-of-the-art predictive method. 

In addition, one of the advantages of PREs is that they can deal with datasets where the number of potential predictor variables is (much) larger than the number of observations. Note, however, that if the number of potential predictor variables becomes much larger, this makes the prediction problem more complex.

\subsection{Computation Time}

As with most statistical methods, the computational burden of fitting PREs increases with the number of observations and/or predictor variables. For (very) large datasets, the default settings of \verb|pre| may yield long computation times. Most of the computational burden is due to the fitting of a large number (500, by default) of decision trees, in order to generate the rules.

Computation time of function \verb|pre| can be reduced in three main ways: Firstly, by employing CART instead of conditional inference trees for rule generation. It should be noted, however, that CART suffers from a variable selection bias towards predictors with a larger number of unique values. This variable selection bias may also present in the fitted PRE and  thus employing CART may only be recommended if all predictor variables are of the same type. Variable selection bias in PREs has not been empirically studied yet, but the results of \citeA{Fokkinpress} suggest that the use of CART yields more complex and often less accurate final ensembles. 

Secondly, a random-forest instead of a boosting approach to rule induction can yield shorter computation times. With package \textbf{pre}, a random-forest approach can be employed by specification of the $mtry$ argument, which will result in a random subsample of $mtry$ predictor variables being selected as potential candidates for every split in every tree. Furthermore, setting the boosting parameter or learning rate to a value of 0 may yield a slight decrease in computation time, as sequential updating of the response variable will then no longer be performed. Adjusting these model-fitting parameters will yield different predictive performance and complexity of the final ensemble, but there is no general rule for determining whether a random-forest or boosting approach will perform better. 

Finally, computation time can be further reduced by employing the parallel computation option of \textbf{pre} for estimation of the final ensemble.

\subsection{Optimizing Model-Fitting Parameter Settings}

The default settings of \textbf{pre} represent the author's choice of parameter settings, which can be expected to perform relatively well for most datasets. The defaults were chosen so as to yield ensembles with relatively low complexity and high predictive performance. These defaults can be adjusted, based on the characteristics of a specific dataset, or the preferences or subject-matter expertise of the researcher. For example, the maximum number of conditions in rules can be specified, based on the researcher's prior expectation on the order of interactions present in the data, or to obtain rules that are easier to interpret and apply. Again, a trade-off between complexity and predictive performance may apply, where increased predictive performance may come at the cost of increased complexity. 

However, the association between complexity and predictive performance of the final ensemble is generally non-linear and will depend on the data problem at hand. Thus, users may often want to determine the optimal values of a set of model-fitting parameters for a specific data problem. \textbf{R} package \textbf{caret} (short for classification and regression training; \citeNP{Kuhn08}) can be used to assess the effect of model-fitting parameter values on predictive performance and complexity of a given algorithm by means cross validation. Package \textbf{pre} allows for the use of \textbf{caret} to optimize the values of a the model-fitting parameters that are most likely to affect predictive performance and complexity of the final PRE. The \textit{Supplementary Material} provides an example showing how the parameter settings of \textbf{pre} can be tuned using \textbf{caret}.

\newpage
\section{Acknowledgements}
The authors would like to thank Benjamin Christoffersen for his contributions to the development of package \textbf{pre}. The authors would like to thank Susan Niessen for granting access to the dataset from Example 2 (Predicting Academic Achievement). Example 3 (Predicting Substance Use) results from secondary analyses of data from a clinical trial conducted by the National Institute on Drug Abuse (NIDA). Specifically, data from NIDA CTN-0044 "Web-delivery of Evidence-Based, Psychosocial Treatment for Substance Use Disorders" were used. NIDA databases and information are available at \url{https://datashare.nida.nih.gov}. Part of the work in this publication was carried out during a research visit of the first author to the University of Zurich, supported through an International Short Visit grant from the Swiss National Science Foundation (IZK0Z1\_175531).\\

\newpage

\bibliographystyle{apacite}
\bibliography{bib}

\end{document}